\documentclass{elsart}

\usepackage{natbib}
\usepackage{graphicx}

\usepackage{amssymb}

\begin{document}

\begin{frontmatter}



\title{Accretion--disc radius variations in close binaries}


\author[jmh]{Jean-Marie Hameury},
\ead{hameury@astro.u-strasbg.fr}
\author[jpl]{Jean-Pierre Lasota}
\ead{lasota@iap.fr}

\address[jmh]{Observatoire de Strasbourg,
67000 Strasbourg, France}
\address[jpl]{Institut d'Astrophysique de Paris--UMR 7095 du CNRS,
75014 Paris, France }

\begin{abstract}
Outer radius variations play an important role disc structure and
evolution. We consider theoretical and observational consequences of
such variations in cataclysmic binaries and low-mass X-ray binaries.
We find that the action of tidal torques must be important well
inside the tidal radius. We also conclude that it is doubtful that
the tidal-thermal instability is responsible for the
superoutburst/superhump phenomena.
\end{abstract}


\end{frontmatter}

\section{Introduction}

Dwarf novae are a subclass of cataclysmic variables undergoing
outbursts lasting (at least) a few days during which their
brightness increases by several magnitudes \citep[see e.g.][for a
review]{w95}. These outbursts are widely believed to be due to a
thermal/viscous disc instability \citep[see][for a review]{l01}. The
weakest point of this model -- aside from the assumption that
angular momentum transport is due to viscosity (i.e. is a local
phenomenon accompanied by energy dissipation) described by the
$\alpha$-prescription, is the approximate treatment of 2D effects at
the disc edge.

There is in particular a debate about the outcome of the disc
reaching the radius at which the 3:1 resonance occurs (this may
happen for low secondary to primary mass-ratios). SPH models
\citep[see e.g.][]{w88} treat accurately the dynamics of the disc
and are in principle quite appropriate to deal with these effects;
they predict that when the 3:1 resonance radius is reached, the disc
becomes tidally unstable, eccentric and precesses. SPH models are,
however, limited in that it is difficult to include in them detailed
microphysics; the energy equation is often replaced by isothermal or
adiabatic approximations; this significantly affects the results as
showed by \citet{kr00}. Nevertheless, the tidal instability, coupled
with the standard thermal instability is also believed \citep{o89}
to be the reason for the long duration of superoutbursts during
which superhumps are seen.

Another point of debate is the amplitude of the torque $T_{\rm
tid}$ due to the tidal forces, even far from the resonance. In a
steady state, the radius of the accretion disc is determined by
$T_{\rm tid}$. One often uses the prescription of \citet{s84},
derived from the linear analysis of \citet{pp77} confirmed by
numerical simulations by \citet{io94} \citep[at least for radii
not too close to the tidal truncation radius $r_{\rm tid}$, at
which the trajectories of test particle orbiting around the white
dwarf intersect, leading to high dissipation][]{p77}. This torque
reads:
\begin{equation}
T_{\rm tid} = \breve{c}{2 \pi \over P} r \nu \Sigma
\left(\frac{r}{a}\right)^n \label{torque}
\end{equation}
where $P$ is the orbital period, $a$ the orbital separation and
$\breve{c}$ a numerical constant. $\nu$ and $\Sigma$ are
respectively the kinematic viscosity coefficient and the surface
density. The index $n$ was found to be close to 5. In the particular
case of U Gem, \citet{io94} obtained a value for $\breve{c}$ smaller
by almost two orders of magnitude than that required for the disc to
be truncated at the tidal radius in steady state. They concluded
that the tidal torque is very small everywhere except very close to
the tidal radius $r_{\rm tid}$ where $T_{\rm tid}$ diverges. If
true, the tidal removal of the angular momentum would occur only at
the disk's outer edge within a negligibly small radial extent
\citep{s02}.

We consider here various aspects of outer disc radius variations. We
first concentrate on disc sizes observed in eclipsing VY Scl
systems. \citep[see][]{hw96}.
We then turn to SU UMa stars
and Soft X-ray Transients (SXTs) and examine for these systems the
consequences of assuming that superhumps and superoutbursts are due
to a tidal instability.

\section{The tidal torque and VY Scl stars}

VY Scl stars are a subgroup of cataclysmic variables which are
usually in a bright state, and have occasionally low states during
which their luminosity drops by more than one magnitude, bringing
them into the dwarf-nova instability strip. Yet, they do not have
dwarf-nova outburst, even though the decline can be very gradual and
prolonged, longer than the disc's viscous time. It had been
suggested that the apparent stability of these systems could be due
to the irradiation of the inner disc edge \citep{lhk99}, but
\citet{hl02} showed that outbursts are unavoidable unless the disc
disappears completely during quiescence, most probably truncated by
the white-dwarf's magnetic field. This translates into a requirement
on the magnetic field strength that must be sufficient for the
magnetospheric radius to be roughly equal to the circularization
radius when the accretion rate is just equal to the critical rate
below which instabilities would appear.

Recently, \citet{skb04} observed the eclipsing VY Scl star DW UMa in
a state intermediate  between minimum and maximum. Eclipse mapping
techniques allowing to reconstruct the disc luminosity profile, they
found that the luminosity difference between the high and
intermediate states is almost entirely due to a change in the
accretion disc radius, from $\sim$ 0.5 to $\sim 0.75$ times $R_{\rm
L_1}$, the distance from the white dwarf to $L_1$. It is observed
that in the intermediate state, the disc is entirely eclipsed by the
secondary, while its outer parts are visible during the high state.
A possible explanation of this could be that the disc is not in
equilibrium (i.e. $\dot{M}$ is not constant within the disc), so
that radius variations are a consequence of mass redistribution
inside the disc. This, however, cannot be the case, since, as noted
by \citet{skb04}, the recovery from the low state takes $\sim$ 4
months; the disc has enough time to readjust its structure to
changes of the mass-transfer rate. The disc should then be quasi
steady, and its radius close to the tidal radius if tidal torques
were exerted only in very small annulus at $r \sim r_{\rm tid}$.
Observations show, however, a disc radius $\sim 50\%$ smaller than
$r_{\rm tid}$ \citep{skb04}.
\begin{figure}
 \center{\includegraphics[width=6cm]{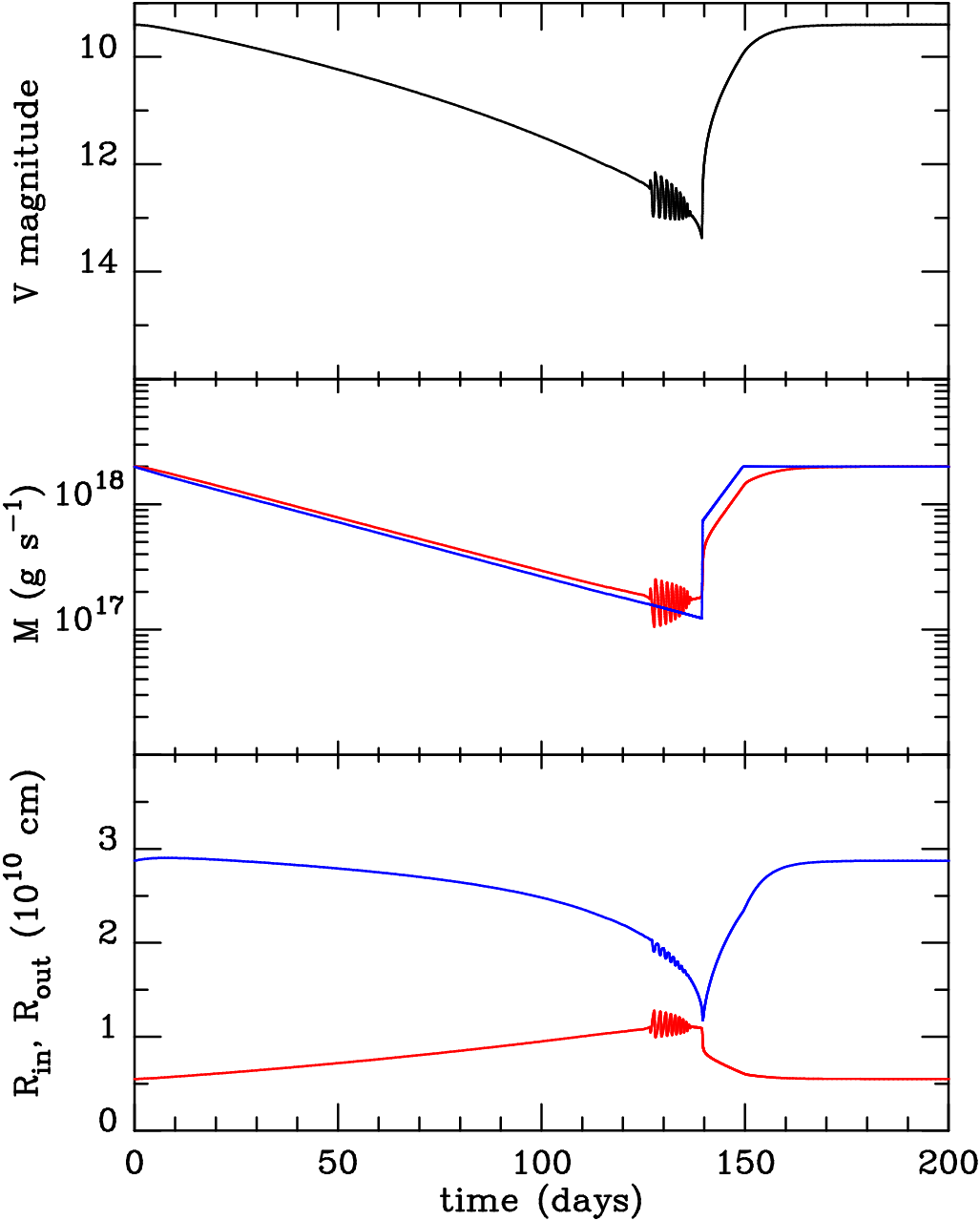}}
 \caption{Evolution of a system with the parameters of DW UMa
 resulting from changes of the mass transfer from the secondary. The
 top panel shows the visual magnitude of the system, the intermediate
 one the mass transfer from the secondary (blue curve) and the mass
 accretion rate onto the white dwarf (red curve, slightly below the
 previous one), and the bottom panel the inner an outer disc radius.
 The magnetic moment is 8 10$^{32}$ G cm$^3$, and the primary mass
 0.7 M$_\odot$ }
 \label{fig:evoldw}
\end{figure}
We used our disc-instability model code
\citep{buat1} to simulate the accretion-disc properties of DW UMa.
We used Eq. (\ref{torque}) with a large $\breve{c}$.

Figure \ref{fig:evoldw} shows the evolution of a system with
parameters similar to those of DW UMa, in which the white dwarf is
magnetized strongly enough to prevent dwarf nova outbursts during
intermediate and low states. There are still some small oscillations
left but these could be suppressed if the field were slightly
stronger, and are probably not detectable. We have assumed a slow
decrease of the mass transfer rate $\dot{M}_{\rm tr}$ until the disc
almost disappears, followed by a rapid (essentially for numerical
reasons) increase of $\dot{M}_{\rm tr}$. The second panel shows that
the disc is always close to equilibrium ($\dot M \approx \dot M_{\rm
tr}$). It can be seen that significant variations of the outer disc
radius are obtained. When the disc is still fainter by 1 magnitude
than the maximum, the disc is 20\% smaller than its maximal
extension. This is not quite the 50\% variations that are claimed by
\citet{skb04}, but is acceptable in view of the fact that we define
the disc outer edge as the place where the surface density vanishes,
whereas \citet{skb04} use a photometric definition.

These results show that the tidal coupling cannot be negligible in
the intermediate state, even at radii quite a bit smaller than the
tidal truncation radius.

\section{SU UMa systems}

As mentioned earlier, the popular explanation for superoutbursts
combines the thermal instability with a tidal instability that is
supposed to arise when the disc reaches the 3:1 resonance radius
\citep[see e.g.][]{o89}. At this point, the disc would become
eccentric and precess, allegedly causing the superhumps. The tidal
torque $T_{\rm tid}$ is supposed to increase by at least one order
of magnitude, resulting in a corresponding enhancement of
dissipation and angular momentum transport. According to this
scenario the disc shrinks until the radius has decreased to an
(arbitrarily chosen) critical value of the order of 0.35 times the
orbital separation, and the superoutburst stops. A sequence of
several normal outbursts, during which the disc grows on average
then follows until the next superoutburst.

This model is essentially based on SPH simulations indicating that
the disc does become eccentric and precesses when the 3:1 resonance
radius is reached \citep{m00,tmw01}, and on the fact that SU UMa
systems are found below the period gap, for systems in which the
mass ratio $q$ is less than 1/3 -- the condition for the 3:1
resonance--radius to be smaller than the tidal truncation radius.
According to simulations by \citet{tmw01} the critical value of $q$
above which no superoutburst occur is in fact rather 1/4.

However, two elements seem to have seriously put in doubt the
viability of the thermal-tidal model of super- humps and outbursts.
First, \citet{kr00} found that the outcome of numerical simulations
depends on the equation of state that has been used; they found that
eulerian models gave the same results as the SPH codes in the
isothermal approximation, but found very different results (no
superhumps) when using the full thermal equation. More recently
\citet{sw05} discovered superhumps in the famous U Gem
1985--superoutburst. The component masses of this prototypical
binary are rather well constrained giving a rather large value of
$q=0.364\pm 0.017$. Clearly the tidal instability cannot apply to
this system. In addition, a permanent superhump was detected in TV
Col \citep{retter03} a binary with an estimated mass-ratio between
$q=0.62$ and 0.93 \citep{hellier93}, consistent with its 5.39 hour
orbital period. One could try to save the tidal model by arguing
that U Gem and TV Col superhumps are phenomena different from those
observed in SU UMa stars but such an argument would have to explain
why, as remarked by \citet{sw05}, when plotted on superhump
excess-period vs orbital period (logarithmic) plane, U Gem and TV
Col fall on the linear extension of the relations defined by shorter
period dwarf novae and permanent superhumpers \citep[see][Fig.
20]{patt03}.

Clearly a new model for SU UMa stars is needed. Irradiation and
enhanced mass transfer probably play a significant role
\citep{hlw00}, but one must find an explanation for the superhump
and superoutburst phenomena keeping in mind that none of them
seems to require the action of tidal forces.

\section{Soft X-ray transients}

In SXTs, the mass ratio is usually very small, and one would
expect to find superhumps in these systems, if indeed the
explanation of this effect is related to the tidal instability.
Indeed there is observational evidence of modulations at periods
slightly longer than the orbital period \citep[][see also Charles,
these proceedings]{z02}. There are, however, several important
differences between SXTs and SU UMa's: first, because the mass
ratios are extremely small, the disc should remain permanently
eccentric, and second, the outer parts of the disc can remain
unaffected by the thermal instability in systems where the disc is
large. Finally, in SXTs superhumps would arise from a modulation
of the reprocessed flux by the changing disc area and not from an
increase of viscous dissipation as in SU UMa stars \citep{hkmc01}.

As an example, for $q = 0.05$, the circularization radius is $0.42
a$, where $a$ is the orbital separation, larger than the critical
radius below which the tidal instability cannot be maintained
\citep[typically $0.35 a$, see e.g.][]{o89}. The disc can therefore
never shrink  enough for the tidal instability to stop, and it
should remain in a permanently eccentric state. This conclusion
holds for values of $q$ below 0.10, i.e. for most SXTs. Therefore
the tidal instability cannot be the cause of SXT outbursts.

\subsection{Short period systems}

For systems with periods $P$ less than $\sim 1$ day, such as A
0620-00, the whole disc is affected by the outburst. One therefore
expects that the superhump modulation, if related to the 3:1 tidal
instability, should be visible in these systems. Systems for which a
superorbital modulation has been claimed \citep[XTE J1118+480, Nova
Musca 1991, GRO J0422+32 and GS 2000+25;][]{z02,oc96} all have $P
\leq 1$ d. However, the disc is \textsl{always} larger than the 3:1
resonance radius as confirmed by the presence of the superhump in
quiescence \citep{z02}, hence outbursts in SXTs cannot be due to
tidal interactions. Interestingly superhumps in quiescence have been
also observed in SU UMa stars \citep{patt95}.

\subsection{Long period systems}

\begin{figure*}
 \begin{center}
 \includegraphics[width=6cm]{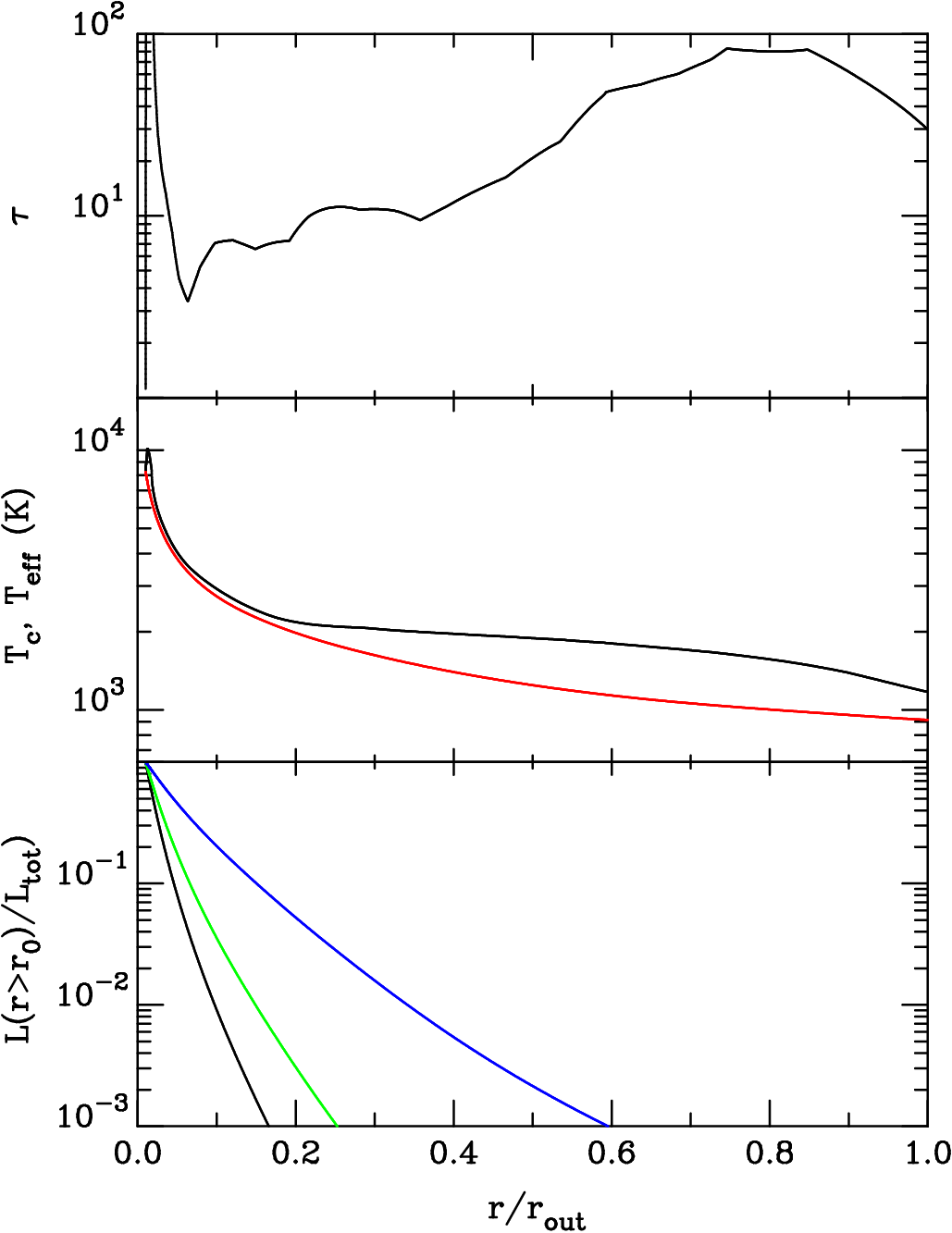}
 \hspace{0.5cm}
 \includegraphics[width=6cm]{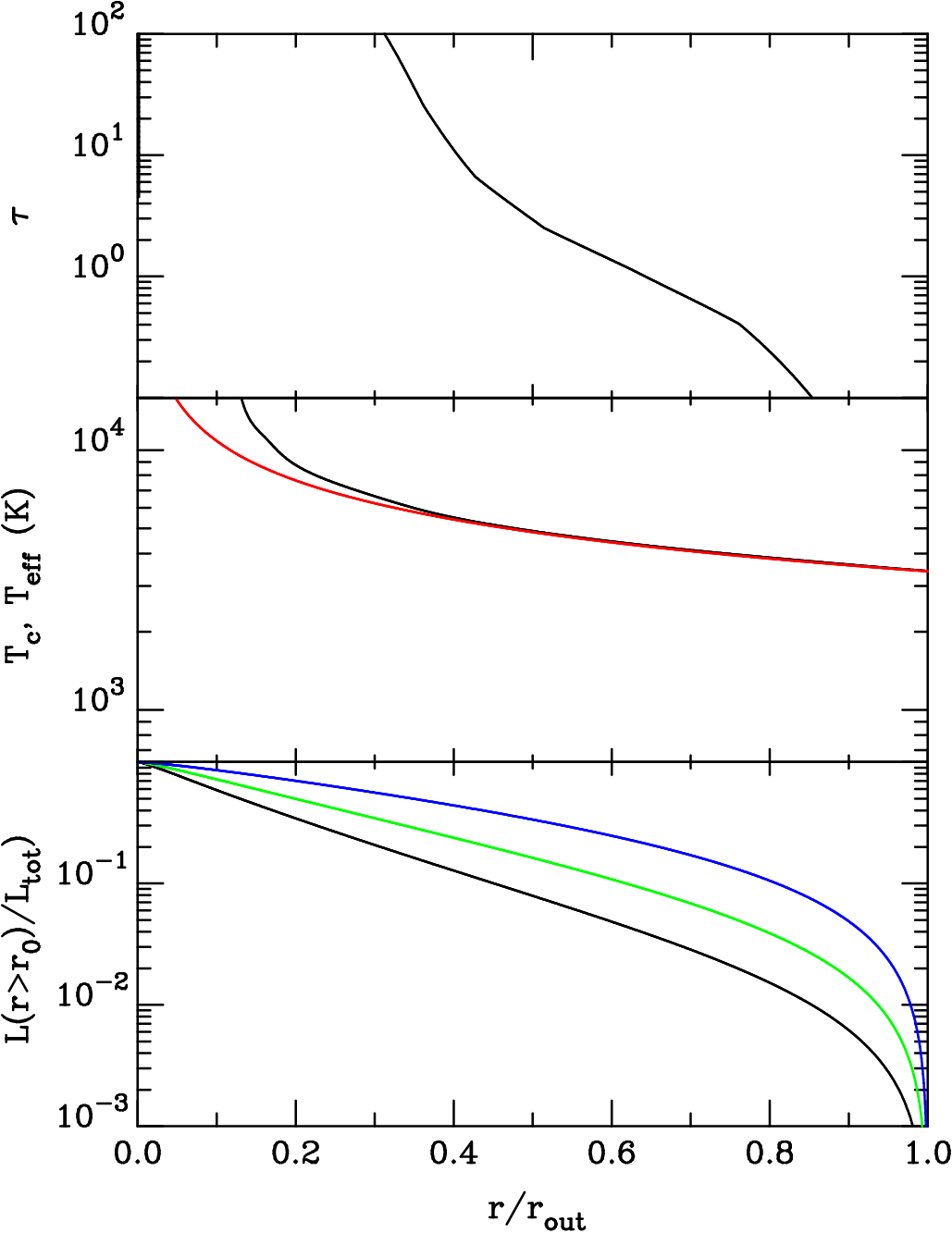}
 \end{center}
 \caption{Structure of the accretion disc in a system with the orbital
 parameters of V 404 Cyg, in quiescence (left), and in outburst (right).
 The top panel shows the optical depth of the disc as a function of
 radius, the intermediate one the central (upper curve) and effective
 (lower curve) temperatures, and the bottom panel gives the fraction
 of the total luminosity emitted above a given radius in the I (upper curve),
 V (intermediate), and B (lower)
 bands}
 \label{fig:sxt}
\end{figure*}

In systems with long orbital periods, the outer disc will not be
affected by the outburst, except possibly by illumination effects.
If the tidal instability sets in, it will therefore always remain
present, and superhumps are expected to be present both in
quiescence or in outburst. It turns out however that in
quiescence, the outer disc is too cold to radiate efficiently even
in the infrared: most of the luminosity originates from the
central regions, despite the reduced emitting area. Figure
\ref{fig:sxt} shows the structure of a disc in a system with the
orbital parameters of black-hole X-ray transient V404 Cyg ($M_1$:
12 M$_\odot$; $M_2$ : 0.7 M$_\odot$; P: 78 hr); the mass transfer
rate was taken to be $10^{16}$ g s$^{-1}$.

In quiescence, the effective temperature is of order of 1000 K,
and less than 0.1 \% of the total luminosity is emitted by the
outer parts of the disc; any modulation of the light emitted by
these cool regions would therefore be undetectable.

In outburst, the disc heats up as a result of irradiation from the
primary (we assumed here the same prescription for irradiation as in
\citet{dlhc99, dhl01}. A significant fraction of the disc luminosity
in the infrared band should therefore be emitted by regions of the
disc possibly affected by the tidal instability. However, the
opacity is minimum for temperatures of the order of 3,000 K, typical
in these regions that are affected by illumination, and it turns out
that the optical depth is very small. As a consequence, the
assumption of blackbody emission does no longer hold, and the colour
temperature will be significantly higher than the effective
temperature; light will be shifted in the optical or blue band, and
will again be diluted in the total light emitted by the inner disc.

We therefore do not expect to detect any superorbital modulation
in these systems, in the hypothesis that it would originate from
the outer parts of the accretion disc (whether it is caused by a
tidal instability or not).

\section{Conclusion}

The effects of the companion on the accretion disc are still
unclear. Even in the simple case where no strong resonance is
present, observations of intermediate states of VY Scl stars
indicate that the tidal torques must be strong enough to truncate
the disc at radii much smaller than the so called tidal truncation
radius.  The presence of superhumps during the superoutburst of U
Gem casts doubt on the validity of the tidal-thermal model of these
phenomena. This conclusion is strengthened by theory and
observations of superhumps in TV Col. If we make the assumption that
superhumps are in some way related to the 3:1 resonance, we predict
that superhumps should be observed in short period SXTs in
quiescence and in outburst, but not in long period systems. The
outburst mechanism should then be unrelated to this resonance, and
the tidal-thermal instability model does not apply to these systems.

\end{document}